\documentclass[a4paper,12pt]{article}

\usepackage{graphicx}

\usepackage{subcaption}

\usepackage[top=2.5cm,bottom=2cm,left=2cm,right=2cm]{geometry}

\usepackage{amsmath}

\usepackage{amssymb}

\usepackage{amsfonts}

\usepackage[affil-it]{authblk}

\usepackage[table,xcdraw]{xcolor}

\usepackage{float}

\everymath{\displaystyle}

\begin{document}

\title{Early- and late-time evolution of Rayleigh-Taylor instability in a finite-sized domain by means of group theory analysis}

\author{ Annie Naveh\\[0.5cm]
Miccal T. Matthews \\[0.5cm]
Snezhana I. Abarzhi\thanks{Corresponding author, email snezhana.abarzhi@gmail.com}}
\affil{School of Mathematics and Statistics, The University of Western Australia, Perth, Western Australia 6009, Australia}

\maketitle

\textit{We have developed a theoretical analysis to systematically study the late-time evolution of the Rayleigh-Taylor instability in a finite-sized spatial domain. The nonlinear dynamics of fluids with similar and contrasting densities are considered for two-dimensional flows driven by sustained acceleration. The flows are periodic in the plane normal to the direction of acceleration and have no external mass sources. Group theory analysis is applied to accurately account for the mode coupling. Asymptotic nonlinear solutions are found to describe the inter-facial dynamics far from and near the boundaries. The influence of the size of the domain on the diagnostic parameters of the flow is identified. In particular, it is shown that in a finite-sized domain the flow is slower compared to the spatially extended case. The direct link between the multiplicity of solutions and the inter-facial shear function is explored. It is suggested that the inter-facial shear function acts as a natural parameter to the family of analytic solutions.}

\section{Introduction}

In 1883 Lord Rayleigh questioned the outcome of supporting a denser layer of fluid  on top of a lighter fluid layer under the effect of gravity, and with this seemly straightforward idea the challenge of understanding the Rayleigh-Taylor instability was born \cite{Rayleigh}. An instability is generally identified as a component within a system that grows without bound. The Rayleigh-Taylor instability (RTI) is a well-known example of an instability in fluid dynamics. It occurs in fluids, plasmas or (in extreme circumstances) materials. RTI develops on the interface of two fluids (or plasmas or solids) of different densities as they are accelerated against their density gradients; that is, the acceleration is directed from the heavy to the light fluid. While the fluid interface is kept perfectly planar and normal to the external acceleration it is in an unstable state of equilibrium and the instability will not form. However, even a small disturbance of this interface is sufficient to initiate the unbounded growth of RTI. For example we observe RTI whenever a denser, heavier fluid such as water is placed on top of a lighter fluid such as oil and the external acceleration in this case is due to gravity. RTI is initiated as potential energy is released and finger-like protrusions are observed as large coherent bubble structures of the lighter fluid penetrates the heavier fluid and spikes of the heavy fluid penetrates the light. A shear function across the interface results in vortical small scale structures (Kelvin-Helmholtz instabilities) on the side of the evolving spikes. This periodic array of bubbles and spikes as the two fluids mix in time is what defines RTI. An exact reliable description of this evolution is the intellectually rich challenge set by Lord Rayleigh decades ago.\\
\\
The challenge to understand the growth of RTI remains relevant today for its role in natural phenomena ranging from astrophysical to microscales; in inertial confinement fusion, laser ablation, combustion, atmospheric flows, as well as in supernovae explosion and the formation of the universe  \cite{ref1laserfusion}\cite{ref1ablation}\cite{ref1atmos}\cite{ref1supernova}. Further the ability to understand and control this instability is vital for industrial applications such as in free-space optical telecommunication, laser micromachining and in aeronautics \cite{ref1laser}\cite{ref1aero}.\\
\\
Previous theoretical studies have considered RTI in an infinitely extended spatial domain in the direction of acceleration. More often however, real world physical phenomena and applications in industry exist in a bounded space. This creates a  need to translate these infinitely extended mathematical models to more useful physical models that may be applied to real world systems. An important question must be asked: how does finite spatial boundaries influence the characteristic evolution of the instabilities? Answering this question not only reveals the particular effects of imposing a  finite domain on the evolution of the instabilities, but also the convergence to the infinitely extended case.

\subsection{Review of approaches} 
 
Lord Rayleigh \cite{Rayleigh} was the first to theoretically study RTI in the incompressible, inviscid fluid case. In doing so, he distinguished himself as not only the first to recognise the significance of hydrodynamic instabilities, but also for his contributions to the theoretical understanding of the initial evolution. Rayleigh found that RTI initially develops faster for smaller values of the wavelength of the initial perturbation, $\lambda$.\\
\\
Since Rayleigh's paper, our understanding of the initial stages of RTI has been enhanced significantly and the idealisations of the inviscid, incompressible fluid overcome. The early time growth rate of the instability has been found for more involved cases taking into account effects such as surface tension, viscosity and compressibility in non-ideal fluids  \cite{lr2}. However, finding a description of the instability evolution in the later, nonlinear time regime presents a more formidable challenge explicit in the long history of the development of solution methods.

\subsubsection{First estimations}

G.I. Taylor \cite{Taylor} was the first to observe the nonlinear evolution of the instabilities experimentally, working with fluids of contrasting densities (specifically water or ethanol and air) in cylindrical tubes. In his work the physical importance of RTI in accelerated fluid layers could first be demonstrated. Taylor considered three-dimensional cylindrical symmetric dynamics in the approximation of a potential flow and driven by a constant acceleration. Taylor numerically estimated a solution to the steady state velocity of RTI and used the tube radius $\lambda/2$ and gravity $g$ as the scales of the problem. He found the velocity of the bubble to be $v\approx0.49 \sqrt{g(\lambda/2)}$ which  corresponded with his experimental results. Indeed he lends his name to the instability for his contributions and predominantly for his cautious experimental observations.

\subsubsection{Layzer-approach and a unique solution}

A major advancement towards a trustworthy description of the nonlinear instability evolution was from Layzer in 1955 \cite{Layzer}. Layzer assumed the flow of a single fluid system to be a potential flow. For incompressible flows a velocity potential satisfies the Laplace equation. Accounting for the conservation laws of mass and momentum, Layzer expanded the potential flow using Bessel functions (for the three-dimensional case) and Fourier series (for the two-dimensional case). He retained only the lowest-order terms (first harmonic) and expanded the conservation laws in the vicinity of the bubble tip.  An important difference of the Layzer approach to the method presented in this work, as part of a group theory approach, is that only a single Fourier harmonic is retained. Although the asymptotic solutions of the Layzer approach was in good agreement with the results of Taylor, it does not account for the interplay of multiple harmonic modes and fails when the harmonic exceeds unity. A multiple harmonic description must be considered as it influences the evolution of the diagnostic parameters.

\subsubsection{Parameter family of solutions}

Garabedian (1957) developed a much deeper analysis that considered multiple harmonics using conformal mapping. He was the first to suggest that there is not one singular steady state solution but a family of solutions governed by a parameter \cite{Garabedian}. This implies that nonlinear solutions are in fact not unique but belong to a finite family bounded by the symmetries of the system. He identified the physically significant solution in the family as the fastest one. This solution, however, differed from the Layzers solution by approximately 10 \%.

\subsubsection{Moments equations}

In 1992 Inogamov advanced the idea of a one parameter family of solutions and introduced moment equations (weighted sums of Fourier amplitudes) to solve the nonlinear system to higher orders. He used potential flow and stream functions to reach a one parameter family of solutions \cite{Inogamov}.

\subsubsection{Group theory}

Results from a group theory approach, first applied to the RTI problem by Abarzhi in 1995 \cite{1995}, ties together and makes sense of the previous results of Layzer (1955), Garabedian (1957) and Inogamov (1992) (all studied for a single fluid system). Using group theory, Abarzhi showed that in the two-dimensional case there is a one parameter family of solutions. In the first-order approximation the fastest solution in the family has a bubble curvature and velocity in agreement with Layzer's solution \cite{1995}. In higher orders the velocity value of the fastest solution increases and agrees with Garabedian's solution. The discrepancy between Layzer and Garabedian's results is due to an order of approximation. Further, there is no Layzer-type solution, only a Layzer first order approximation. The necessity of considering higher-order interactions for obtaining a reliable description is one of the indications of the non-local character of the instability evolution.\\
\\
Finally, the group theory approach finds a stability region for the curvature of the bubble tip (the family parameter) and the corresponding velocity \cite{2008review}. Inogamov's solution sits outside of this stability region. The results obtained from a group theory approach are confirmed in advanced numerical simulations and in more accurate experiments \cite{experimental1} \cite{experimental2}.

\subsection{Previous results of group theory for an infinite domain}

An analysis based on group theory is applicable for many RTI-related problems in two and three dimensions with various symmetries for $0 < A \leq 1$, where $A$ is the system's Atwood number \cite{grouptheory1} \cite{grouptheory2} \cite{2008b} \cite{OG}. Local properties of the evolution are accounted for in local spatial expansions and global properties are accounted for in Fourier series expansions \cite{patternform}. A group theory approach reliably accounts for the asymptotic dynamics for nonlinear RTI bubbles in an infinite spatial domain. A stability analysis suggests that the obtained RTI families are complete as they involve all the possible solutions allowed by the spatial symmetry of the flow \cite{2008review}. For asymptotic time $t/\tau \rightarrow \infty$ (where $\tau$ is a suitable time scale) to the first-order RTI dynamics approaches a steady-state bubble velocity and curvature \cite{finitedensity}.\\
\\
For contrasting fluid densities, $A=1$, and at the first-order of approximation the fastest solution in the parameter family of solutions corresponds with a Layzer-approach approximation. This yields a maximum velocity and corresponding curvature of
\begin{equation}
v_A=\sqrt{\frac{g}{3k}}\quad\text{and}\quad\zeta_A=-\frac{k}{6}
\end{equation} 
where $k$ is the wave number of the Fourier expansion. The critical bubble curvature, $\zeta_{cr}$, is defined as the maximum curvature supported by the system. Independent of Atwood number it is found in an infinite spatial domain as $\zeta_{cr}=-k/2$ and a bubble curvature outside of the region $0\leq|\zeta_1|\leq|\zeta_{cr}|$ is not supported by the convergence of the solutions.

\section{Theoretical modelling approach}

\subsection{System description}

Fluid flow problems in general are very difficult and often impossible to solve analytically without appropriate assumptions set in place. This is due to the equations governing the fluid flow; namely the celebrated Euler or (compressible) Navier-Stokes equations. The set of equations span three dimensions, are time dependent, have pressure gradient terms and include viscous stress. Indeed the specific fluid problem of modelling the nonlinear evolution of acceleration-driven hydrodynamic instabilities is undeniably complex \cite{2008review}. As in many fluid flow descriptions, to successfully navigate this problem and obtain a rigorous description of the diagnostic parameters certain theoretical approximations must be set in place.

\subsubsection{Fluid idealisation}

Firstly, in this work the effects of compressibility, viscous stress and surface tension are neglected. This approximation proves important as finding the proper choice of boundary conditions for compressible or miscible fluids is a very difficult and, to a large extent, an unsolved problem \cite{2008review}.

\subsubsection{Two-dimensional}

\begin{figure}[H]
	\centering
	\includegraphics[width=\linewidth]{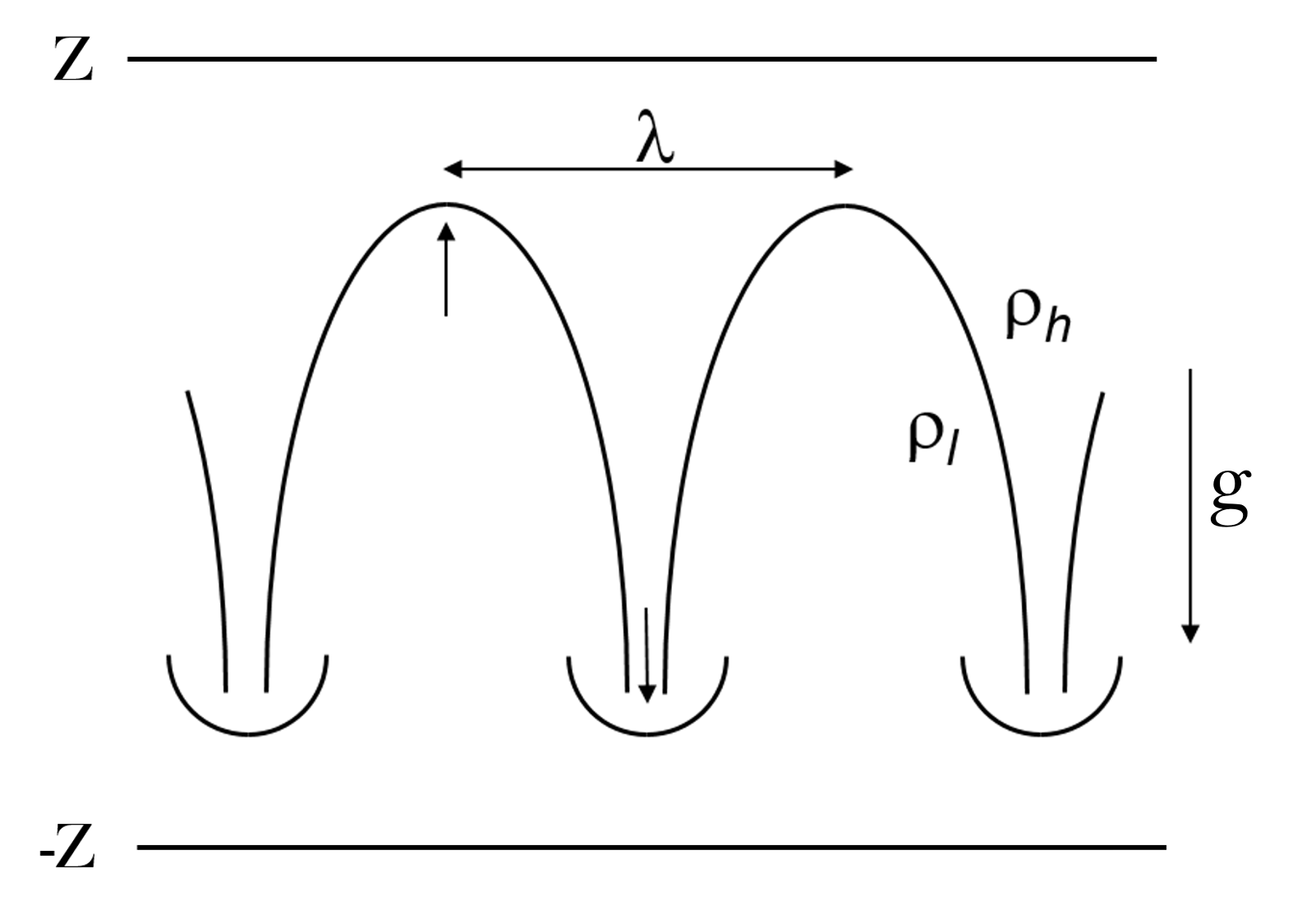}
	\caption{Large-scale coherent structure of bubbles and spikes. $\rho_h$, $\rho_l$ represent the density of the heavy and light fluid respectively. $\lambda$ is the spatial period set by the initial perturbation and $\boldsymbol{g}$ is the acceleration. Arrows mark the direction of fluid motion at the tip of the bubble (up) and spike (down). Imposed at a height $z = Z$ and $z = -Z$ are boundaries restricting the fluid flow (not to scale). For the spatially extended case the boundary position is set as $Z=\infty$ and for the spatially bounded case $Z<\infty$.}
	\label{fig:diagram}
\end{figure}

The flow is assumed to be two-dimensional; periodic in the $x-$direction, motionless in the $y-$direction and spatially bounded in the $z-$direction. The latter is set as the direction of acceleration. Fixed boundaries are imposed for all values of $x$ at positions $z=\pm Z$. The set-up of the problem is illustrated in Figure \ref{fig:diagram}. The extended periodicity of the coherent structures in the $x-$direction is an important symmetry property enabling a group theory consideration. It is for this reason that the finite boundaries are imposed only in the direction of acceleration and the periodic structures repeat infinitely in the $x-$direction. To expedite the calculations, an analysis is performed in the non-inertial frame of reference moving with velocity $v(t)$ in the $z-$direction, where $v(t)$ is the bubble velocity at the bubble tip in the laboratory frame of reference. Solutions are translated back into the laboratory reference frame where necessary.

\subsubsection{Foundational scales}

The evolution of the instability should be found such that it is true at all scales, requiring an identification of the appropriate scales of the problem. The spatial period of the initial perturbation $\lambda$, $\lbrack \lambda \rbrack=m $ and the uniform acceleration $g$, $\lbrack g \rbrack=ms^{-2}$ are the foundational scales for RTI. The length scale may be represented as the wave number $k$ where $k=2\pi/\lambda$ and $\lbrack k \rbrack = m^{-1}$. These basic scales define the time scale of the flow $\tau\sim\sqrt{gk}^{-1}$, the initial bubble growth rate $v_0 \sim \sqrt{g / k}$ and the frequency $\omega\sim\sqrt{g k}$. For a realistic system of non-ideal fluids the characteristic length scale $\lambda$ is set by the mode of fastest growth, whereas for this analysis, regarding ideal fluids, the spatial period $\lambda$ is set by the initial perturbation.

\subsubsection{Scale separation}

For a large density ratio of the two fluids the nonlinear regime dynamics of the RT flow can be categorised into two relatively independent scales \cite{2003}. The large scale set by the initial perturbation wavelength ($\sim\lambda$) includes the coherent structures of bubbles and spikes, while the small-scale set by the amplitude of the initial perturbation ($\ll\lambda$) consists of the vortical structures caused by the shear function. It should be noted that this theoretical separation of scales is not applicable for fluids with very close densities (that is, for small Atwood numbers $A$). This is because as the difference in fluid densities approach zero the vortical structures become large and the scale separation is no longer valid. The small scale vortical structures are a consequence of the non-linearities and secondary instabilities from the full Navier-Stokes equations \cite{multiscale}. These small scale dynamics and the interaction between the scales results in a randomness that convolutes the mixing as a complex process. Fortunately, the multi-scale mixing process does maintain certain features of coherence and order associated primarily with the dynamics of the large scales  \cite{2008b}. To theoretically solve this problem we therefore focus our solutions to the large-scale coherent dynamics; the bubble and spike structures.

\subsubsection{Localised to the bubble-tip}

The mushroom-type spike gets its shape from the non-deterministic small-scale vortical structures. Further to describe RTI evolution in the vicinity of the spike requires a very high order of approximation to deal with the many singularities \cite{2008b}. Therefore to obtain a trustworthy deterministic description of the unstable dynamics in the large scales we find solutions in the vicinity of the coherent, large scale bubble front.

\subsubsection{Time regimes}

To date an accurate theoretical description of the evolution of RTI for all times has not been found \cite{2008review}. Instead studies of these instabilities (and the work presented here) are predominantly concerned with the evolution in early time $t\ll\tau$ (the linear regime) and latter time in which the bubble and spike structures are formed (the nonlinear regime). For a description in the nonlinear regime the dynamical system is solved for asymptotically large time $t/\tau \rightarrow \infty$. The evolution in this regime reveals the true intellectual richness of this problem and so is the main focus of this work.

\subsubsection{Multiple harmonics}

A multiple harmonic analysis of the system is considered such that waves that are integer multiples of the initial perturbation wavelength are retained in our analysis. This is crucial as these higher-order harmonics contribute to the diagnostic parameters of the motion and should be accounted for in the description of the evolution \cite{Inogamov}. The major assumption of the multiple harmonic analysis is that the dynamics of the flow are governed by a dominant mode. For a realistic system of non-ideal fluids the dominant mode and characteristic length scale is set by the mode of fastest growth \cite{finitedensity}. On the other hand, for this analysis regarding ideal fluids, the dominant mode is set by the initial perturbation for the RTI.

\subsection{Governing equations}

The dynamics of the idealised fluid system considered is governed by the set of incompressible Euler equations; a simplification of the more general Navier-Stokes equations with zero viscosity and a constant density imposed. The set of incompressible Euler equations depict, respectively, the conservation of mass, momentum and energy:
\begin{equation} \label{eq:NS}
\frac{\partial \rho}{\partial t} +  \nabla \boldsymbol{\cdot}\left(\rho \boldsymbol{v}\right)= 0 $$$$
\frac{\partial \boldsymbol{v}}{\partial t} +  \left(\boldsymbol{v}\boldsymbol{\cdot} \nabla\right)\boldsymbol{v} +\frac{\nabla P}{\rho} = 0 $$$$
\frac{\partial E}{\partial t} +  \nabla \boldsymbol{\cdot}\left(E+P\right) \boldsymbol{v}= 0
\end{equation}

Here $\rho,\boldsymbol{v},P,E$ is the field of density, velocity, pressure and energy and $t$ is time. The energy field can be further expressed as  $E=\rho \left(e+\frac{\boldsymbol{v}^2}{2}\right)$ with $e$ being a specific internal energy. All the variables are considered to be continuous functions of the spatial coordinates and time. For an ideal incompressible fluid the density and the internal energy of the fluid are constant in the fluid's bulk.

\subsubsection{Inter-facial boundary conditions}

For incompressible, immiscible fluids, the fluid interface of the two fluids is discontinuous and this motivates the introduction of a local scalar function $\theta(x,y,z)$ defined such that $\theta=0$ at the interface, the heavy (denser) fluid is located at $\theta>0$ and the light (less denser) fluid at $\theta<0$. It is assumed that the derivatives $\dot{\theta}$ and $\nabla\theta$ exist. Using this simple function the system can be explicitly expressed as 

\begin{equation}
\left(\rho,\boldsymbol{v},P,E\right)= \left(\rho,\boldsymbol{v},P,E\right)_hH(\theta)+\left(\rho,\boldsymbol{v},P,E\right)_lH(-\theta)
\end{equation} 

where $H(x)$ is the Heaviside function and the subscripts $h$ and $l$ refer to the observables in the heavy and light fluid, respectively. Describing the incompressible two fluid system holistically using the $\theta$ function and substituting it into the governing Euler equations (\ref{eq:NS}) yields the following inter-facial boundary conditions in the case of zero mass flux across the interface
\begin{equation}
\label{eq:intbc}
\rho\left(\frac{1}{\nabla \theta}\frac{\partial \theta}{\partial t}+\boldsymbol{v}\boldsymbol{\cdot}\boldsymbol{n}\right)=0$$$$
\big[\boldsymbol{v}\boldsymbol{\cdot}\boldsymbol{n}\big]=0\quad,\quad\big[\boldsymbol{v}\boldsymbol{\cdot}\boldsymbol{\tau}\big]=\textrm{arbitrary}\quad,\quad
\lbrack P\rbrack=0\quad,\quad\lbrack W \rbrack=\textrm{arbitrary} 
\end{equation} 
The parenthesis $\lbrack\dots\rbrack$ denotes a jump of functions across the interface, and $W=e+\frac{P}{\rho}$ is the specific enthalpy. The inter-facial boundary conditions indicate, respectively, that across the interface; mass flux is continuous, normal component of velocity of the fluid is continuous, tangential component of velocity is \emph{not} continuous and pressure is continuous. As the fluids are considered incompressible we can omit the discontinuity of specific enthalpy $W$ from our consideration. Extra attention should be given to the boundary condition indicating the discontinuity of the tangential component of velocity as this allows a shear function to develop along the interface. Later it is shown that this shear function may act as the natural parameter to the family of solutions.\\
\\
The boundary conditions at the outside boundaries are simply
\begin{equation}
\mathbf{v}_h=0\quad\textrm{at}\quad z=Z\quad,\quad \mathbf{v}_l=0\quad\textrm{at}\quad z=-Z
\end{equation}

\subsubsection{Large-scale dynamics}

Any vector field can be written as the sum of a scalar potential function and a curl and so the velocity vector fluid of the fluids can be expressed as
\begin{equation}
\boldsymbol{v}=\nabla\Phi +\nabla \times \boldsymbol{\phi}
\end{equation}
The large-scale dynamics are assumed to be irrotational therefore $\nabla \times \boldsymbol{\phi}=0$. Indeed in the small scale the flow is not irrotational (due to the interfacial vortical structures). Considering only the large scales and expressing the velocity as a scalar potential is a valid and necessary approximation allowing us to reach a solution analytically. For large scales, therefore, the fluid velocity field is conservative and may be presented by a gradient potential function
\begin{equation}
\label{eq:potential}
\boldsymbol{u}=\nabla\Phi
\end{equation}
for some scalar potential $\Phi$. Limiting our analysis to the large-scale dynamics, the governing Euler equations (\ref{eq:NS})  and derived inter-facial boundary conditions (\ref{eq:intbc}) can be expressed in terms of the scalar potential $\Phi$.\\
\\
Substituting (\ref{eq:potential}) into the conservation equations (\ref{eq:NS}) reduce to
\begin{equation}
\label{eq:Laplace}
\nabla^2\Phi=0$$$$
\rho\left(\frac{\partial \Phi}{\partial t}+\frac{\nabla\Phi^2}{2}\right)+P=0
\end{equation}
Notice that this system is governed by the Laplace equation; this is as expected as the flow is both irrotational and incompressible. To complement the governing conservation equations at large scales the inter-facial boundary conditions (\ref{eq:intbc}) should likewise be expressed in terms of the scalar potential $\Phi$. Substituting (\ref{eq:potential}) into (\ref{eq:intbc}) yields the appropriate inter-facial boundary conditions at large scales given below in the non-inertial reference frame of the bubble-tip:     
\begin{equation}
\label{eq:intbcfull} 
\rho_h\left(\nabla\Phi_h\boldsymbol{\cdot}\boldsymbol{n}+\frac{\dot{\theta}}{|\nabla\theta|}\right)=\rho_l\left(\nabla\Phi_l\boldsymbol{\cdot}\boldsymbol{n}+\frac{\dot{\theta}}{|\nabla\theta|}\right)=0$$$$
\nabla\Phi_h\boldsymbol{\cdot}\boldsymbol{\tau}-\nabla\Phi_l\boldsymbol{\cdot}\boldsymbol{\tau}=\textrm{arbitrary}$$$$
\rho_h\left[\frac{\partial \Phi_h}{\partial t}+\frac{|\nabla\Phi_h|^2}{2}+\left(g+\frac{dv}{dt}\right)z\right]=\rho_l\left[\frac{\partial \Phi_l}{\partial t}+\frac{|\nabla\Phi_l|^2}{2}+\left(g+\frac{dv}{dt}\right)z\right]
\end{equation}
The set of defining equations is completed by defining the outside boundary conditions for the instability in a finite domain
\begin{equation} \label{eq:outsideboundaryfin}
\left.\frac{\partial \Phi_h}{\partial z}\right\vert_{z=Z}=-v(t) \quad,\quad \left.\frac{\partial \Phi_l}{\partial z}\right\vert_{z=-Z}=-v(t)
\end{equation}

\subsection{Method of solution}

\subsubsection{Symmetry groups}

Symmetry in mathematics implies invariance of a given system under certain transformations. The periodicity of the large scale coherent motion in the $x-$direction is a natural property of two-dimensional RTI, and is used as the defining symmetry element. This natural periodic structure is invariant with respect to one of the seven one-dimensional crystallographic symmetry groups. Explicitly these seven groups are $p1$, $p1m1$, $p11\texttt{g}$, $p2$, $p2m\texttt{g}$, $p11m$, and $p2mm$ using international classification and Fedorov's notation. However, not all of the seven invariant groups should be considered as an appropriate portrayal of RTI. This is because the instability flows are essentially anisotropic and the dynamics in the $z-$direction of acceleration differs from that in the other directions. An additional requirement for the correct symmetry group is one with coherent structures that are observable and repeatable -- we need to find a group such that the symmetry properties do not change over time and therefore are structurally stable. This is satisfied if the group is a symmorphic group with inversion in the plane \cite{coherent}. Group $p1m1$, hereafter $pm1$, generators are translation $x + \lambda \rightarrow x$  and mirror reflection in the $(z,y)$ plane  $ x \rightarrow -x$ and is the symmetry group most suitable for the analysis of the coherent structures of RTI growth.

\subsubsection{Fourier series expansion}

Using a group theory approach with a strong consideration for the symmetries in the system inspires the application of a Fourier Series in solving the governing Laplace equation. In the case of continuous translational symmetry  $x + \lambda \rightarrow x$ in which the translation parameter $\lambda$ can take any value, the set $\sin{(kx)}$ forms a complete set of irreducible representations of the odd functions and $\cos{(kx)}$ for the even functions. The Fourier series are hence irreducible representations of the group of translations. For the $pm1$ group the fluid potentials are expressed as

\begin{equation}
\Phi_h =\sum_{m=1}^{\infty}\Phi_m(t)\left[ z \sinh(mkZ) +\frac{\cos(mkx)}{mk}\cosh\left[mk(z-Z)\right]\right]+f_h(t)
$$$$
\Phi_l =\sum_{m=1}^{\infty}\tilde{\Phi}_m(t)\left[-z \sinh(mkZ)+\frac{\cos(mkx)}{mk}\cosh\left[mk(z+Z)\right]\right]+f_l(t)
\end{equation}
where $\Phi_m(t)$ and $\tilde{\Phi}_m(t)$ are, respectively, the Fourier amplitudes of the heavy and light fluid corresponding with the mode $m$, and $f_h(t)$ and $f_l(t)$ are time-dependent functions.

\subsubsection{Spatial expansion}

In the vicinity of the bubble tip the interface, set as $z^*$, evolution must take into account the $pm1$ generator $ x \rightarrow -x$, simplifying the Taylor expansion of the bubble tip to
\begin{equation}z^*(x,t)=\sum_{i=0}^{N}\zeta_i(t)x^{2i}\end{equation} where $N$ is the order of approximation. In this work the case of $N=1$ will be examined due to the complexity of the problem. In the frame of reference of the bubble tip $\zeta_0=0$ and so to the leading order ($N=1$); $z^*(x,t)=\zeta_1(t) x^2$. As the bubble tip takes a predominantly quadratic form $\zeta_1(t)$ is the principal curvature of the bubble tip and as the bubble is concave we look for solutions corresponding to $\zeta_1(t)<0$. \\
\\
For this expansion to be valid we assume the following conditions $\frac{x}{\lambda}<<1$ and $\frac{z-z^*}{Z}<<1$ or equivalently for the last constraint $z-z^*<<Z$. Hence the expansions and following solutions hold only for a sufficiently large domain. Solutions are being considered for a \emph{finite but large} domain.

\subsubsection{Moments}

The weighted sums of the infinite number of Fourier amplitudes are named the \emph{moments} and expressed as
\begin{equation}
\begin{split}
M_n=\sum_{m=1}^{\infty}\Phi_m(t)(mk)^n\sinh(mkZ)\\ \tilde{M}_n=\sum_{m=1}^{\infty}\tilde{\Phi}_m(t)(mk)^n\sinh(mkZ)\\N_n=\sum_{m=1}^{\infty}\Phi_m(t)(mk)^n\cosh(mkZ)\\ \tilde{N}_n=\sum_{m=1}^{\infty}\tilde{\Phi}_m(t)(mk)^n\cosh(mkZ)
\end{split}
\end{equation}
where $n=0,1,2\dots$\\ 
\\
Using these moment expressions the dynamical system in $N=1$ is derived for the first time to give the conditions for the continuity of mass flux, normal component of velocity and pressure in a finite domain at the interface and at the outside boundaries of the finite domain as
\begin{equation}
\label{eq:dynamfinRT}
\dot{\zeta_1}-3N_1\zeta_1-\frac{M_2}{2}=
\dot{\zeta_1}-3\tilde{N}_1\zeta_1+\frac{\tilde{M}_2}{2}=0$$
$$(1+A)\left(\frac{\dot{N_1}}{2}+\zeta_1\dot{M_0}-\frac{N_1^2}{2}-g\zeta_1\right)=(1-A)\left(\frac{\dot{\tilde{N_1}}}{2}-\zeta_1\dot{\tilde{M_0}}-\frac{\tilde{N}_1^2}{2}-g\zeta_1\right)
$$$$
M_0=-v(t)\quad,\quad\tilde{M}_0=v(t)\quad,\quad M_0=-\tilde{M}_0
\end{equation} 
and the condition for the discontinuity of the tangential component at the interface is
\begin{equation}
N_1-\tilde{N}_1=\textrm{arbitrary} 
\end{equation}
Recall that $A$ is the Atwood number of the two fluid system and defined as $A=\frac{\rho_h-\rho_l}{\rho_h+\rho_l}$ and a dot marks a time-derivative. \\
\\
Expressing the system in terms of infinite sums provides the principal opportunity to derive the regular asymptotic solutions for $t/\tau \rightarrow\infty$ with a desired accuracy that accounts for the higher-order spatial modes and enables a stability analysis of the asymptotic dynamics. To find the solutions for the dynamical system in the linear and nonlinear regimes, we have to solve the closure problem. The closure problem can be addressed using the arguments of symmetry; specifically by considering all the local asymptotic solutions allowed by the spatial symmetry of the flow. For the discrete group $pm1$ the number of family parameters $N_p=1$. The expressions for the moments in the defining equations should be expanded such that in every order of approximation $N$, the number of equations $N_e$ and the number of variables $N_t$ (the Fourier amplitudes $\Phi_m,\tilde{\Phi}_m$ and surface variables $\zeta_i$) obey the relation $N_t\geq N_e+N_p$. In this case, $N_p=1$ and $N_e=4$ indicating that $N_t\geq5$. To find the nonlinear asymptotic solution the proper order of approximation of the moments should be taken to solve the closure problem and establish the proper relations between the moments.

\section{Results}

The diagnostic parameters of RTI i.e the growth rate and curvature of the bubble tip, can be characterised for early and asymptotically late time. These time regimes are named, respectfully, the linear and nonlinear regime. The strength of the outlined methodology is that it is applicable in both these time regimes and for systems with a range of Atwood numbers. For consistency solutions are given for systems of all Atwood numbers $ 0 < A\leq1$. It should be stressed, however, that the potential separation from which the equations are derived are only valid for large Atwood numbers and so we should proceed cautiously when approaching $A \rightarrow 0$. Further these solutions hold only for a \emph{finite but large} domain size. Although for completion the solutions are given in a form that appears to hold for all boundary heights including $kZ\rightarrow 0$, in this small boundary limit the results should be received tentatively. \\
\\
Prior to discussing the specific effects of imposing a finite domain on the growth of the instability it is useful to explore the convergence of the finite domain to the infinitely extended case as $kZ\rightarrow\infty$. This convergence is found to be independent of Atwood number and for any fixed curvature a critical scaled boundary height of $kZ=5$ is established. At this domain size the percentage difference between the diagnostic parameters in the finite domain and those in the infinite domain falls below a negligible 0.02 \%. In other words, at a boundary height of $kZ=5$ the effects of the finite boundaries on the instability become negligible and therefore is a suitable approximation for an infinite domain.

\subsection{Linear regime -- unique solution}

For early time $t/\tau\ll1$, namely the linear regime, the perturbation amplitude is small and all harmonics above the first order are disregarded
\begin{equation}
\Phi_m(t)\quad,\quad\tilde{\Phi}_m=0\quad\forall \quad m>1
\end{equation}
as it is assumed that for early time the multiple modes have not had time to form. In the linear time regime we work under the approximation of an initially small bubble curvature and perturbation amplitude, $0<\vert \zeta_1(t_0)k \vert<<1$ and $0<\vert(k/g)^{1/2}v(t_0)\vert<<1$ where $t_0$ represents the initial instant of time and $v(t_0)$ the initial growth rate. For early time $t\approx t_0$ and $v\approx |v(t_0)|$, therefore enabling a trustworthy simplification of the governing dynamical system to the first-order in $v(t_0)$ and $t_0$ by assuming that the contribution of higher orders is negligible. The governing system (\ref{eq:dynamfinRT}), retained to the first order in $v(t)$ and $t$ is reduced to a set of simplified coupled ordinary differential equations 
\begin{equation}
\label{eq:linearfin}
\dot{\zeta_1}+\frac{v k^2}{2}=0
$$ $$
\dot{v}k+2A g\zeta_1\tanh(kZ)=0
\end{equation} 
Solving these equations yields the growth and curvature rate of the initial RTI perturbation at the fluid interface. We find that for RTI bounded at $Z$ and $-Z$ in the direction of acceleration, the early time perturbation amplitude and curvature in the case of constant acceleration increases exponentially with time $t$ at a rate 
\begin{equation}
v(t)\sim\exp\left(t\sqrt{\tanh(kZ)}/\tau\right)
$$ $$
|\zeta_1(t)|\sim\exp\left(t\sqrt{\tanh(kZ})/\tau\right)
\end{equation} 
where $\tau=\sqrt{Agk}^{-1}$ is the characteristic time scale. Introducing a finite domain decelerates the exponential growth of the perturbation amplitude for early time by an exponent of $\sqrt{\tanh(kZ)}$. Recall that $kZ$ and $-kZ$ are the positions of the imposed boundaries in the direction of acceleration (given in a dimensionless scaled form). The smaller the domain the greater this exponential growth is hindered. In the limit $kZ\rightarrow\infty$ this result agrees with the growth in an infinitely extended spatial domain \cite{OG}. 

\subsection{Nonlinear regime -- family of solutions}

Identifying the evolution of RTI in the nonlinear regime is a more involved task than that in the linear regime. In RTI with constant acceleration, as $t/\tau\rightarrow\infty$, the velocity, moments and surface variables asymptotically approach steady values to the leading-order in time, with the next order terms decaying exponentially with time \cite{2003}. To reflect the steady-state conditions as $t/\tau\rightarrow \infty$ the time derivatives for both the moments and the surface curvature variable, $\zeta_1$, in the dynamical system (\ref{eq:dynamfinRT}) are set to zero.\\
\\
The next consideration in solving the dynamical system (\ref{eq:dynamfinRT}) in the nonlinear regime is correctly satisfying the closure requirement. Truncating the moment expressions such that $\Phi_m, \tilde{\Phi}_m =0$ $\forall$ $m>2$ and for an order of approximation in the spatial expansion $N=1$, retains the variables $\Phi_1, \tilde{\Phi_1}, \Phi_2, \tilde{\Phi}_2 \text{ and } \zeta_1$. In this case the number of variables $N_t = 5$ and sufficiently overcomes the closure problem. Therefore, to obtain a reliable description of the large scale coherent dynamics of the bubble front that correctly captures the influences of higher harmonics the dynamical system in (\ref{eq:dynamfinRT}) is truncated to two harmonic modes. \\
\\
Solving this dynamic system finds a family of all possible bubble velocity solutions parametrised by curvature. The solution family is complete and finds all the solutions supported by the prescribed global symmetries.\\ 
\\
The analytic expression is
\begin{equation}
v=-\frac{12 \sqrt{-Ag\zeta_{1} } \sinh ^2(k Z) \left[2 \zeta_{1}  \sinh (k Z)+k \cosh (k Z)\right]}{k^{2} \left[3 \sinh (k Z)+\sinh (3 k Z)\right] \sqrt{\frac{A \left(k^2-4 \zeta_{1} ^2\right)+A \left(4 \zeta_{1} ^2+k^2\right) \cosh (2 k Z)-4 \zeta_{1}  k \sinh (2 k Z)}{\left[k \cosh (k Z)-2 \zeta_{1}  \sinh (k Z)\right]^2}}}
\end{equation}

\begin{figure}[H]
	\centering
	\begin{subfigure}{0.49\textwidth}
		{\includegraphics[width=\linewidth]{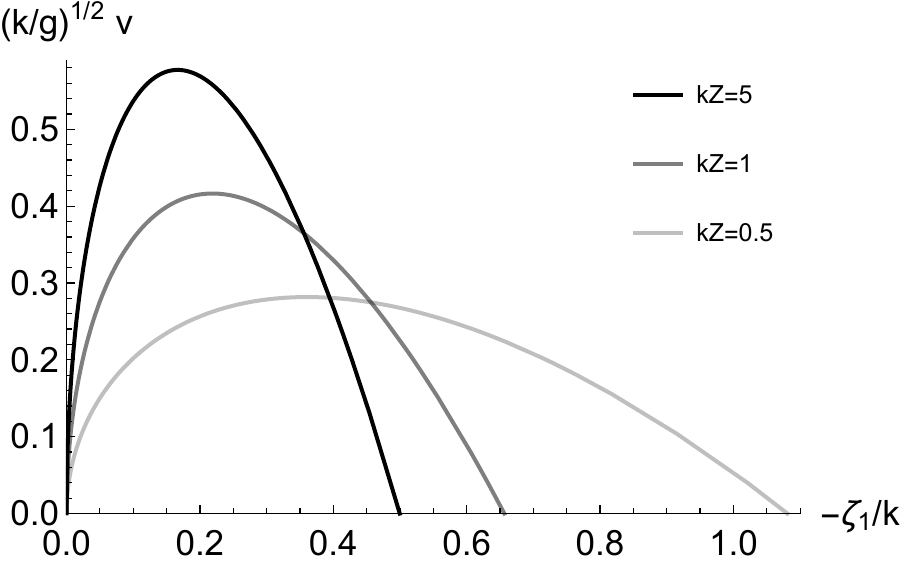}}
		\caption{ }
	\end{subfigure}
	\begin{subfigure}{0.49\textwidth}
		{\includegraphics[width=\linewidth]{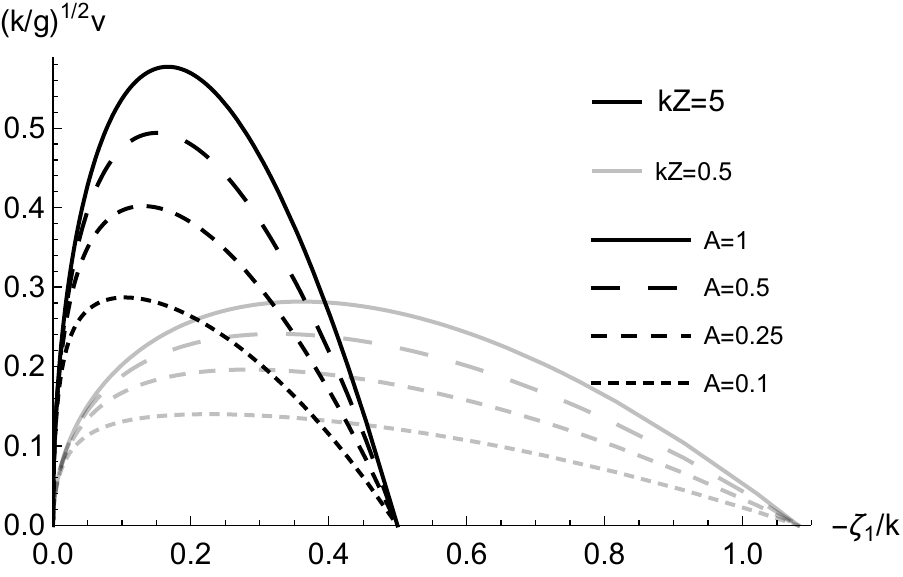}}
		\caption{ }
	\end{subfigure}
	\caption{Influence of a finite domain on the diagnostic parameters of Rayleigh-Taylor instability for a fluid system with (a) $A=1$ and (b) a range of Atwood numbers. Decreasing the boundary flattens the family of solutions vertically and stretches it horizontally; \emph{uniformly} for each Atwood number.}
	\label{fig:fincurve}
\end{figure}

Figure \ref{fig:fincurve} illustrates the family of bubble velocity solutions for various sized finite domains of height $kZ$.\\
\\
The solution family is plotted in dimensionless, positive units. Recall that the bubble curvature, $\zeta_1$, is a negative value, as the bubble is concave, and constant in time in the nonlinear regime for RTI. The units of velocity come from the fundamental velocity scale $\sqrt{g/k}$. \\
\\
Two fundamental consequences of imposing a finite domain on the dynamic parameters of RTI are clearly illustrated. In a finite domain the maximum velocity solution is decreased and the greatest amount of curvature supported by the symmetries of the system is increased. These two responses are enhanced the smaller the finite domain imposed. This means that compared to RTI's evolving in an infinite unbounded spatial domain, RTI bubble fronts growing in a finite domain travel slower with the same bubble front curvature. This result that the boundary hinders motion is physically intuitive. \\
\\
Travelling at the same fixed velocity, RTI bubble structures are more curved in a finite domain than in an infinite domain. Indeed the maximum bubble front curvature supported by the symmetries of the system, denoted as $\zeta_{cr}$, is a monotone decreasing function of the boundary size:
\begin{equation}
\label{eq:critcurve}
\left|\frac{\zeta_{cr}}{k}\right|=\left|\frac{\coth (kZ)}{2}\right|
\end{equation}
Note that for $kZ\rightarrow\infty$, $\zeta_{cr}=-k/2$ and for $kZ\rightarrow 1$, $\zeta_{cr} \approx -0.657k$. Figure \ref{fig:fincurve}b shows the effect of a finite boundary on a range of Atwood numbers. We find that the family parameter of solutions is altered uniformly for all Atwood numbers. The values for $kZ=1/2$ are given only for the purposes of completeness.

\subsection{Atwood bubble}

\begin{figure}[H]
	\centering
	\begin{subfigure}{0.49\textwidth}
		{\includegraphics[width=\linewidth]{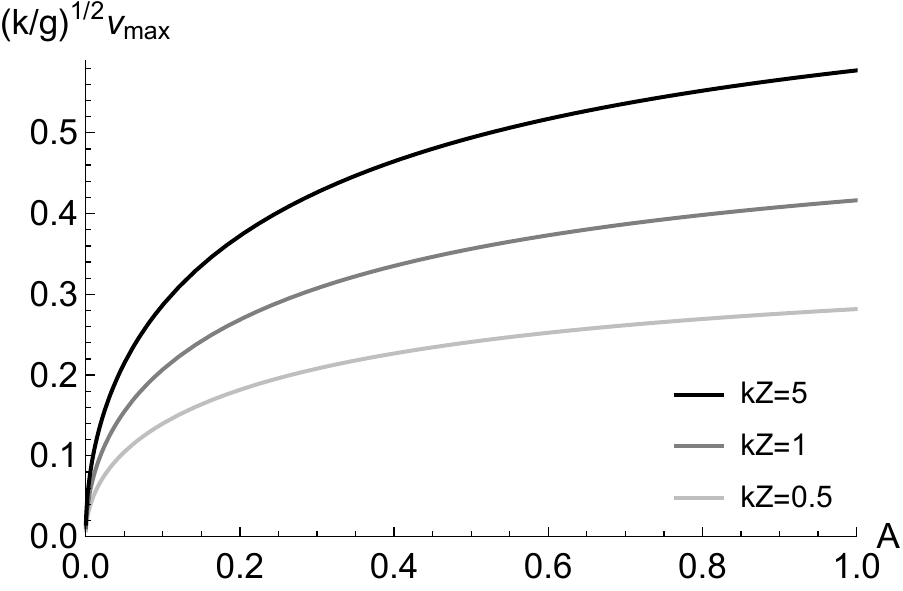}}
		\caption{}
	\end{subfigure}
	\begin{subfigure}{0.49\textwidth}
		{\includegraphics[width=\linewidth]{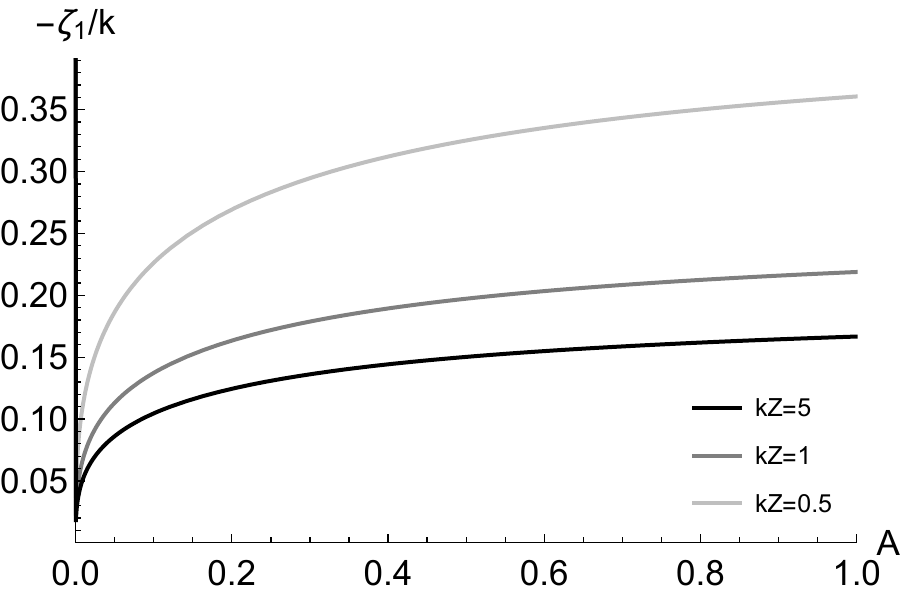}}
		\caption{}
	\end{subfigure}
	\caption{Observables for the fastest Rayleigh-Taylor bubble solution given as a function of Atwood number. (a) Maximum velocity solutions and (b) the curvature corresponding to the fastest solution.}
	\label{fig:vmax}
\end{figure}

\begin{figure}[H]
	\centering
	\begin{subfigure}{0.49\textwidth}
		{\includegraphics[width=\linewidth]{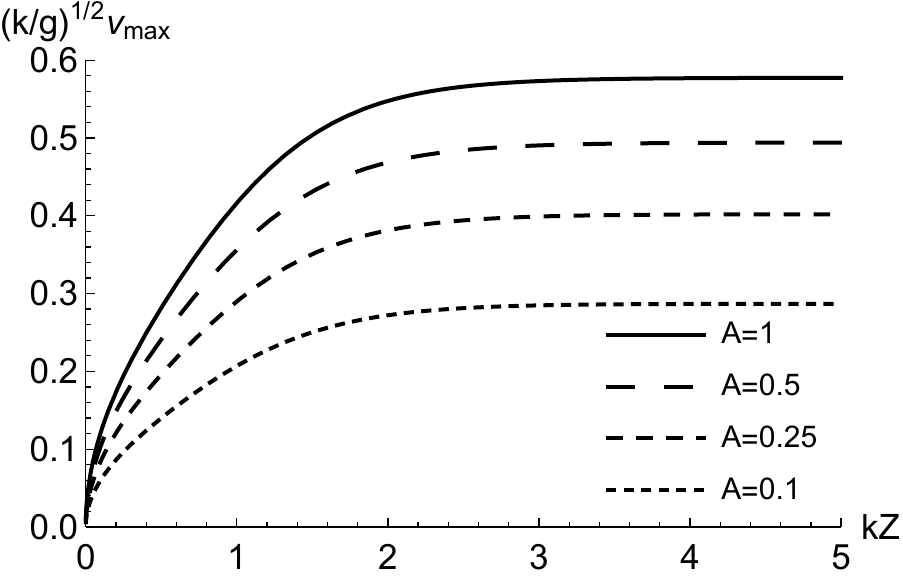}}
		\caption{}
	\end{subfigure}
	\begin{subfigure}{0.49\textwidth}
		{\includegraphics[width=\linewidth]{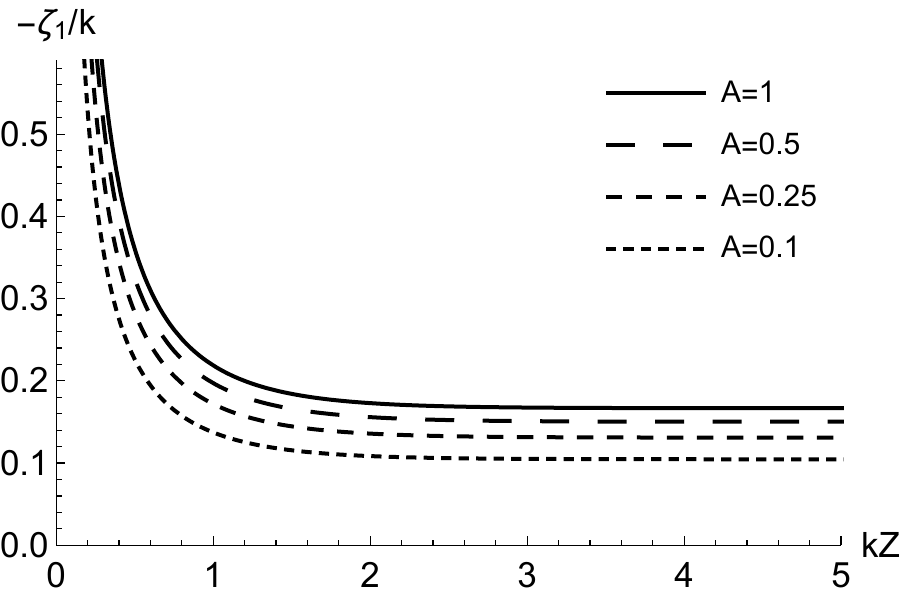}}
		\caption{}
	\end{subfigure}
	\caption{Diagnostic parameters (a) velocity and (b) curvature for the fastest Rayleigh-Taylor bubble in the family of solutions given as a function of domain size.}
	\label{fig:maxvel}
\end{figure}

The parameter family of solutions can be represented by the Atwood bubble. The Atwood bubble is the fastest solution in the family with velocity $v_A$ and corresponding curvature $\zeta_A$ such that 
\begin{equation}
\left.\frac{\partial v}{\partial \zeta_1}\right|_{\zeta_1=\zeta_A}=0 \quad \text{and} \quad \left.\frac{\partial^2 v}{\partial \zeta_1^2}\right|_{\zeta_1=\zeta_A}<0
\end{equation}
Notice that $\zeta_A$ is the curvature corresponding to the maximum velocity solution, not the maximum curvature solution which is instead denoted $\zeta_{cr}$. The curvature of the Atwood bubble $\zeta_A$ satisfies 
\begin{equation}
3AC^4+8C^3+6AC^2-A=0\quad\Rightarrow\quad C=-\frac{2 \zeta_A}{k}\left(\frac{2 \tanh{(kZ)}-\tanh(2kZ)}{\tanh(kZ)\tanh(2kZ)}\right)
\end{equation} 
and $C>0$. The system is described using hyperbolic tangents, for convenience $T1$ and $T2$ are henceforth used as shorthand notation for $\tanh(kZ)$ and $\tanh(2kZ)$, respectively. \\
\\
Transformation scalings are introduced to bridge between solutions in a finite and infinitely extended spatial domain. The fundamental scales of a fluid system in a finite domain may be expressed as a function of the infinitely extended domain. 
\begin{equation}
k=k_{scale}k^*=\frac{T1T2}{2T1-T2}k^*$$ $$g=g_{scale}g^*=\frac{(T1T2)^4}{(2T1-T2)^2(T1-2T2)^2}g^*
\end{equation} 
where the superscript $^*$ corresponds to the parameter in the infinitely extended spatial domain. These fundamental scales lead to the following additional transformation terms:
\begin{equation}
v_{scale}=\sqrt{\frac{g_{scale}}{k_{scale}}}=\frac{(T1T2)^{\frac{3}{2}}}{(2T1-T2)^{\frac{1}{2}}(2T2-T1)}$$ $$\tau_{scale}=\frac{1}{\sqrt{k_{scale}{g_{scale}}}}=\frac{(2T1-T2)^{\frac{3}{2}}(2T2-T1)}{(T1T2)^{\frac{5}{2}}}$$ $$\omega_{scale}=\frac{1}{\tau_{scale}}\quad, \quad \lambda_{scale}=\frac{2\pi}{k_{scale}}
\end{equation}
The solution governing the local dynamics of the bubble front has a non-trivial dependence on the Atwood number and the size of the domain. The solution retains the first four harmonics but for all values the lowest order harmonics are dominant. The general expressions for $v_A$ and $\zeta_A$ are cumbersome and hence not presented here. The simplest information to be drawn from these cumbersome expressions is the existence of an invariant property satisfying
\begin{equation}
\left|\frac{k^3v_A^2}{g\zeta_A^3}\right|=-\frac{72 (2 T1-T2)^2}{(T1-2 T2)^2}
\end{equation} 
and holding for all Atwood numbers. This is similar to what is found in the infinite domain case \cite{OG}. Dependency of the Atwood bubble on $A$ for various boundary sizes is illustrated in Figure \ref{fig:vmax}.\\
\\
For any fixed Atwood number RTI bubble structures travel faster and with a less curved bubble front as the domain size increases. Figure \ref{fig:maxvel} demonstrates the influence of a finite domain on the diagnostic parameters as a function of the boundary size $kZ$. For any fixed boundary size, the RTI grows faster and with a more curved bubble front for contrasting fluid densities. Convergence is seen for both Figures \ref{fig:vmax} and \ref{fig:maxvel} and as the domain size is increased the diagnostic parameters coincide with the infinitely extended case. For similar fluid densities, $A\approx0$, it is well demonstrated that the bubble velocity and curvature are forced to zero tainting the accuracy of the solutions in that vicinity.\\
\\
Analytic expressions for the Atwood bubble may be simplified in the limiting cases of $A\approx 1$ and $A\approx 0$ truncated to give the first two harmonics. The case of $A\approx 1$ is a natural consideration and the theory is applicable with no limitations. Alternatively the properties of the system for $A\approx 0$ is presented for completeness but it is noted that the solutions may be out of range of applicability of the theory.\\
\\
For highly contrasting fluids, $A\approx 1$, we express the diagnostic parameters to the first-order as
\begin{equation}
\zeta_A\approx-\frac{k\left[8-(1-A)\right]}{48}k_{scale}
$$ $$ 
v_A\approx\frac{\left[16-3(1-A)\right]\sqrt{g}}{16\sqrt{3k}}v_{scale}
\end{equation}

For similar density fluids, $A\approx 0$, the fastest solution variables are explicitly
\begin{equation}
\zeta_A\approx-\frac{kA^{1/3}}{4}k_{scale} \quad, \quad v_A\approx\frac{3\sqrt{g A}}{2\sqrt{2k}}v_{scale}
\end{equation}

\subsection{Taylor's bubble}

Taylor’s bubble is named here as the particular solution for the Atwood bubble in a one fluid system, $A = 1$. In this case the problem variables have the form
\begin{equation}
\zeta_T=-\frac{k}{6}k_{scale}\quad, \quad v_T=\sqrt{\frac{g}{3k}}v_{scale}
\end{equation}
Notice that due to the $(-T1+T2)$ term, the second harmonic of the heavier fluid goes to zero $\Phi_2\rightarrow 0$ as the boundary height, and hence domain size, increases $Z\rightarrow\infty$.

\subsection{Qualitative velocity fields}

\begin{figure}[H]
	\begin{subfigure}{0.49\textwidth}
		{\includegraphics[width=\linewidth]{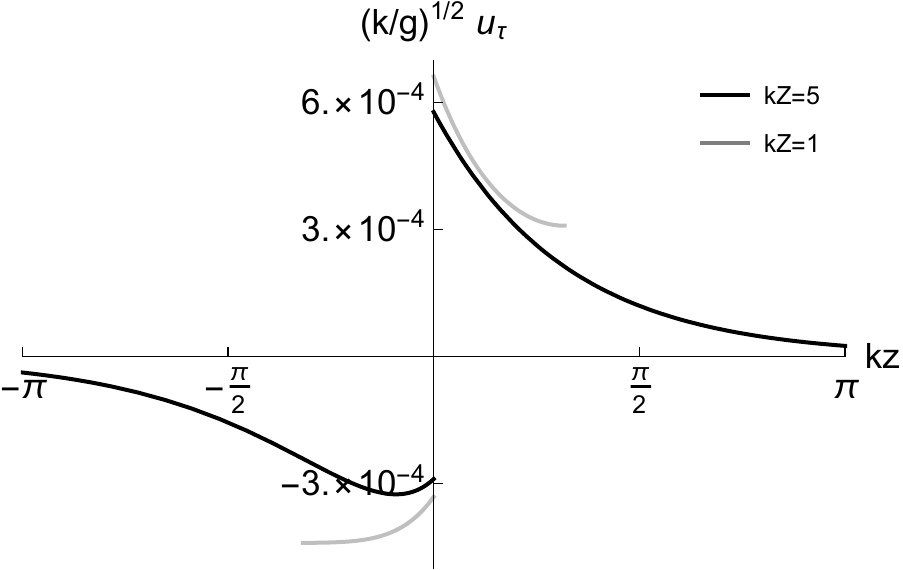}}
		\caption{}
	\end{subfigure}
	\begin{subfigure}{0.49\textwidth}
		{\includegraphics[width=\linewidth]{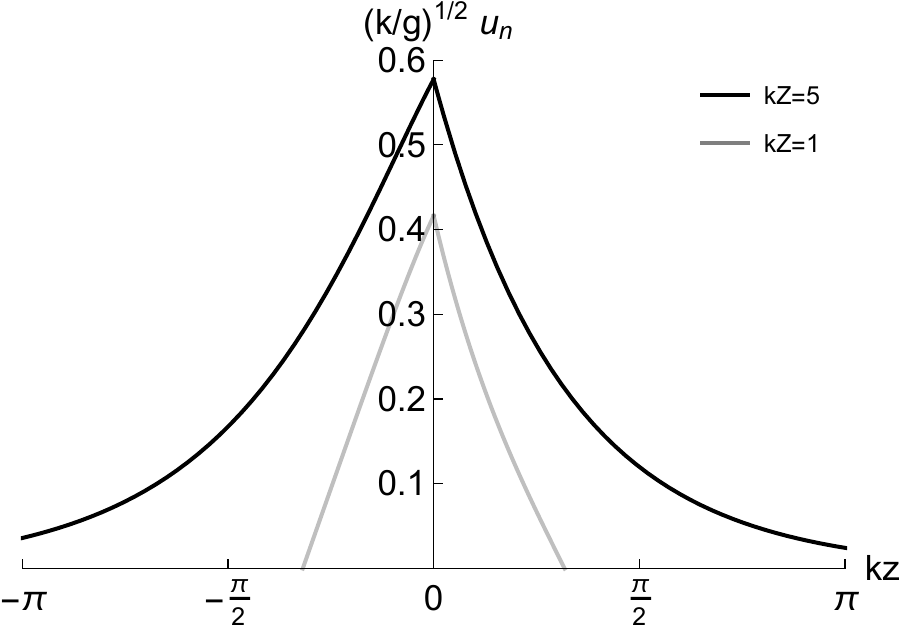}}
		\caption{}
	\end{subfigure}
		\caption{(a) Tangential and (b) normal components of Rayleigh-Taylor instability velocity field for the fastest solution at $kx = 10^{-3}$ and $A=1$. }
	\label{fig:ncaZRTall}
\end{figure}

\begin{figure}[H]
	\centering
	\begin{subfigure}{0.49\textwidth}
		\includegraphics[width=\linewidth]{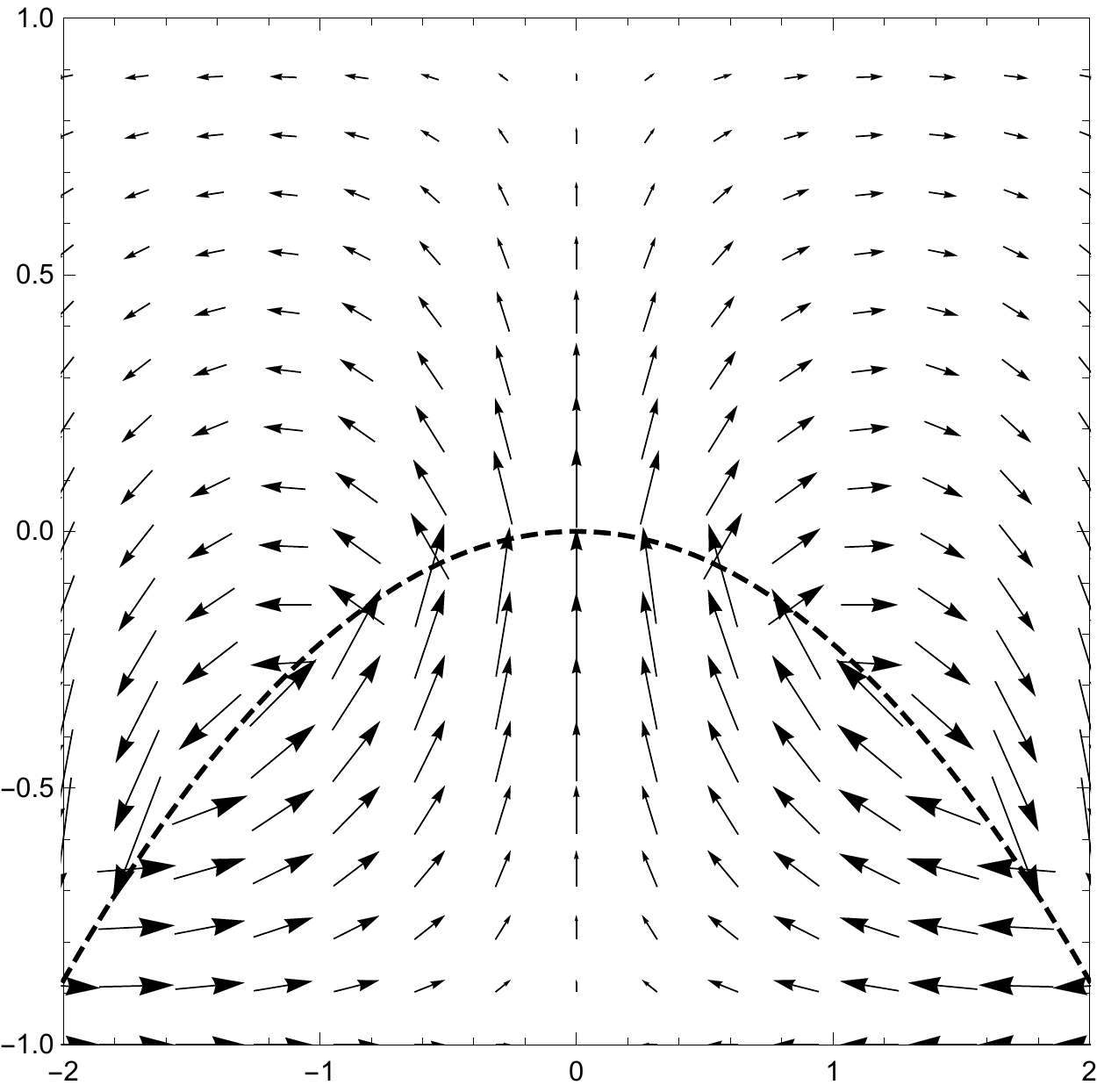}
		\caption{}
	\end{subfigure}
	\begin{subfigure}{0.49\textwidth}
		\includegraphics[width=\linewidth]{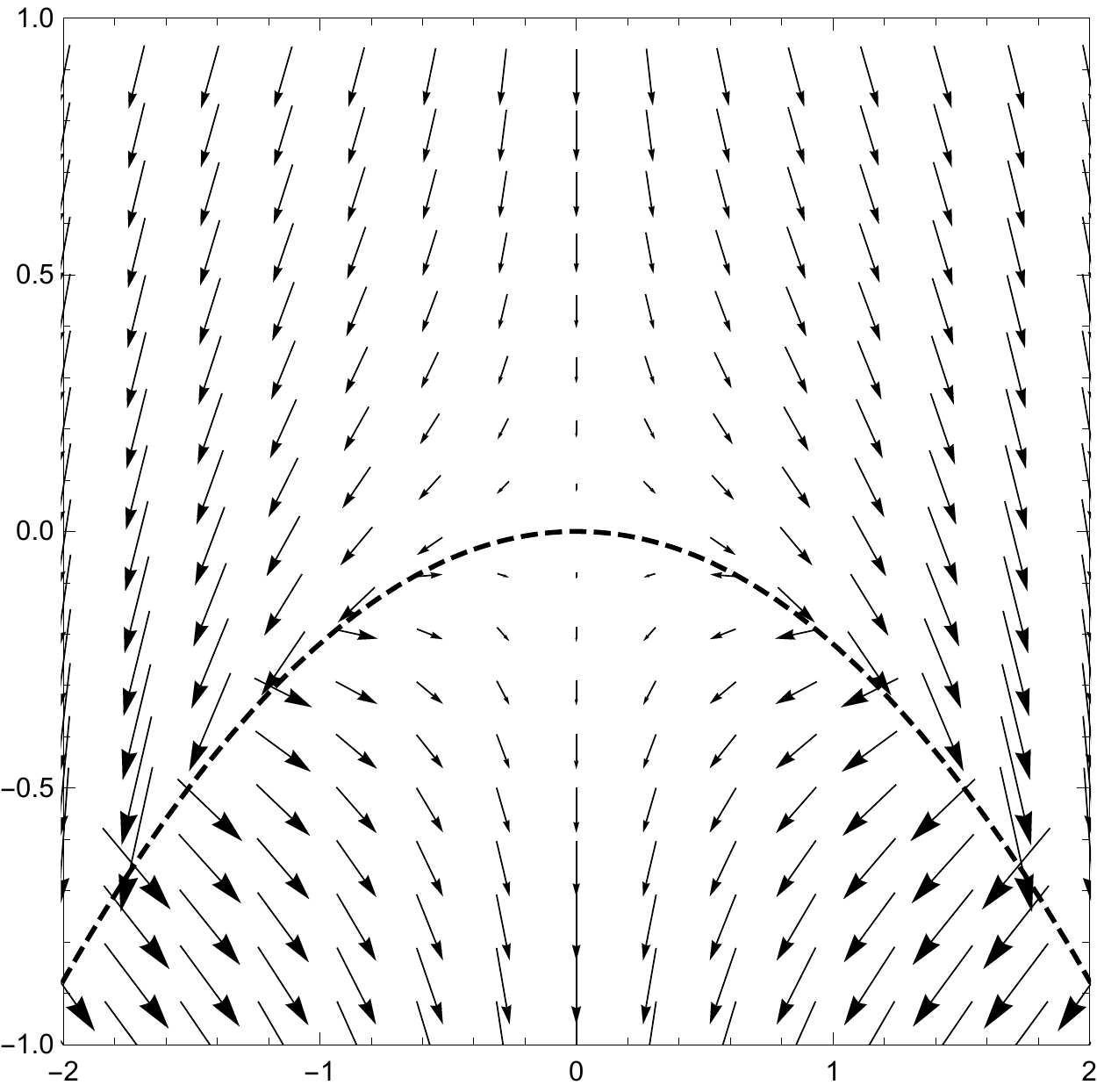}
		\caption{}
	\end{subfigure}
	\caption{Velocity field of Rayleigh-Taylor in a finite boundary $kZ=1$ with respect to (a) the laboratory frame of reference and (b) the non-inertial bubble tip frame of reference. Plotted is the fastest solution with $A=1$ and corresponding curvature is $\approx -0.22k$.}
	\label{fig:vectorplotZ1RTlab}
\end{figure}

In the approximation of a potential flow the velocity of the bubble is given by 
\begin{equation}
\nabla \Phi=\left(\frac{\partial\Phi}{\partial x},0,\frac{\partial\Phi}{\partial z}\right)
\end{equation}
Taking the partial derivatives of the potential function, $\frac{\partial\Phi}{\partial x}$ and $\frac{\partial\Phi}{\partial z}$, we may examine the tangential ($x-$direction) and normal ($z-$direction) components of the velocity as a function of $kz$, the dimensionless displacement from the interface. Here we consider only small distances from the interface, $z\approx z^*$. The system is described such that the heavy (light) fluid is situated at $kz>0$ ($kz<0$). This enables a specific examination of the influence of a finite boundary on the vector components of the fluid motion, see Figure \ref{fig:ncaZRTall}.\\
\\
The results are given in the inertial, laboratory frame of reference and the physically significant (fastest) solution in the family at $A=1$ is examined.  We find that the normal component of velocity is greatest and continuous at the interface. Indeed, this continuity condition was imposed as one of the interface conditions, that is, the condition of no mass flux across the interface. The tangential component of velocity is discontinuous, implying that a shear function develops at the interface. This presence of a shear function leads to the Kelvin-Helmholtz small scale vortical structures. The value of the tangential component of velocity in Figure \ref{fig:ncaZRTall}a is low at a short horizontal distance from the bubble tip, $kx=10^{-3}$. For both tangential and normal components, anisotropy is observed in the direction of acceleration as the velocity profile is asymmetric about the origin. We observe that this anisotropy is enhanced in a smaller finite domain. Further enforcing a smaller spatial domain increases the magnitude of the tangential component of velocity while the normal component decreases. This indicates that the shear function is greater (and hence we can expect larger vortical structures) in RTI evolving in a smaller, finite domain when compared to a larger or infinite spatial domain.\\
\\
Figure \ref{fig:vectorplotZ1RTlab} shows the qualitative velocity fields of RT flows both in the laboratory and non-inertial bubble tip frame of reference. The dashed line represents the interface of the fluids. As described in the system description, the heavy fluid layer is located above the interface and the light fluid below. The constant externally imposed acceleration is downwards, from heavy to light.\\
\\
In the laboratory frame of reference, Figure \ref{fig:vectorplotZ1RTlab}a, RTI is characterised by intense motion of the fluids in the vicinity of the interface and effectively no motion away from the interface. A shear function is present at the interface and is greater further from the bubble tip which may cause vortical structures. There is no shear function exactly at the bubble tip, $x=0$. The method of solution applied to the large-scale dynamics of the instabilities finds the large scale `envelope' properties of the small scale vortical structures \cite{OG}. According to the velocity field in Figure \ref{fig:vectorplotZ1RTlab}a the vortical structures rotate `inwards from light to heavy'.\\
\\
In a finite domain the fastest solution travels with a greater curvature and the inter-facial vortical structures appear more pronounced when compared to the infinite domain evolution \cite{OG}, in agreement with the results of Figure \ref{fig:ncaZRTall}a. At the upper and lower boundaries the normal component of velocity vanishes and only a tangential component remains as set by the boundary conditions
\\
This velocity field agrees qualitatively with accurate experiments and simulations and suggests that linear and nonlinear RT dynamics in incompressible immiscible fluids (1) is essentially inter-facial, with intense motion of the fluids near the interface and effectively no motion of the fluid away from the interface; (2) has potential flow in the bulk and vortical structures at the interface. Our theoretical results are valid in the vicinity of the bubble tip. To fully describe the mushroom-type shape of the spike, further investigations are required, including the non-ideal effects to be done in the future.
\\
\section{Shear function analysis}

Previous studies of RTI consider a bubble curvature parametrisation to the family of solutions. This means that the diagnostic parameters are given as a function of curvature. A parametrisation of curvature is favoured as it is observable and hence comparisons with experiments are simplified. A closer analysis, however, indicates that the shear function may be a more natural and suitable parameter to the solutions. The shear function is defined here as a quantification of the discontinuity of the tangential component of velocity. Indeed, it is this discontinuity that is responsible for the multiplicity of solutions and is believed to significantly influence the nonlinear dynamics \cite{OG}.  Mathematically the shear function, $\Gamma$, is defined as  
\begin{equation} 
\Gamma=\lim_{x\to0}\frac{v_h(x,0)-v_l(x,0)}{x}
\end{equation} 
where $v_h(x,0)$ and $v_l(x,0)$ are the tangential components of the velocity field in a vicinity of the interface between the heavy and light fluid, respectively. The shear function is an important dynamical parameter as it drives the Kelvin-Helmholtz vortical structures. The theoretical approach used in this work allows the quantification of the shear function at the interface in the vicinity of the bubble tip. Although the methodology limits our analysis to the large-scale dynamics, considering the shear function enables us to determine the large scale qualitative properties of the vortical structures such as the relative strength and rotation direction. In the linear regime there is no shear function in the vicinity of perturbation front, this result fits with observations as the vortical structures are not visible for early time. \\
\\
In the nonlinear regime the shear function is clearly present due to the noticeable Kelvin-Helmholtz vortical structures. To leading-order in terms of the moment equations for RTI the shear function is given, in terms of moments, as
\begin{equation}
\label{eq:sheardef}
\Gamma=\tilde{N_1}-N_1
\end{equation}

\subsection{Shear function as a diagnostic parameter}

\begin{figure}[H]
	\centering
	\begin{subfigure}{0.49\textwidth}
		\includegraphics[width=\linewidth]{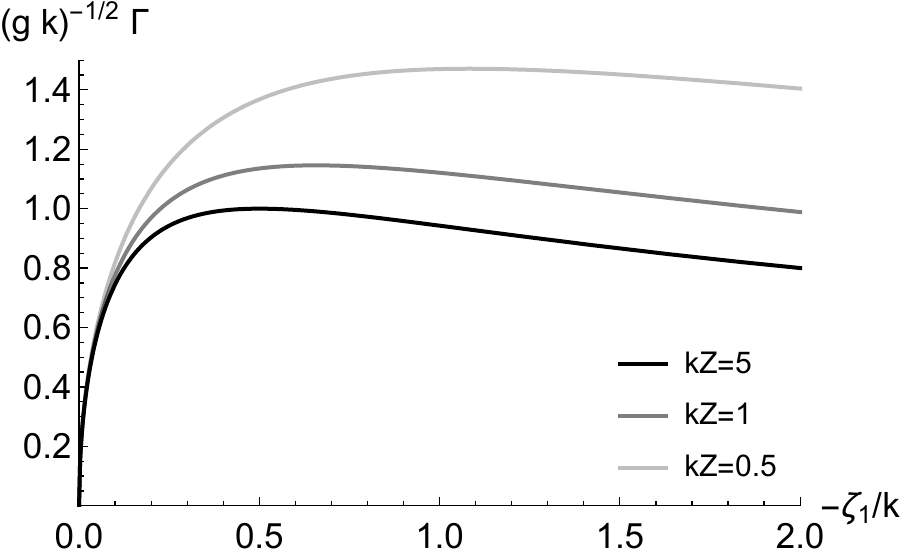}
		\caption{}
	\end{subfigure}
	\begin{subfigure}{0.49\textwidth}
		\includegraphics[width=\linewidth]{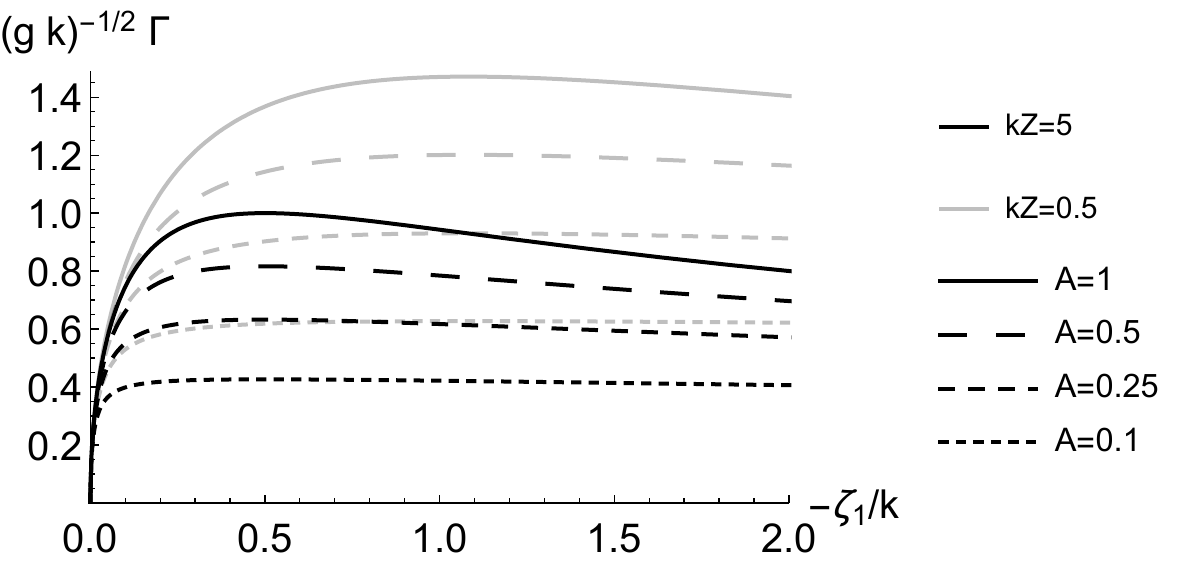}
		\caption{}
	\end{subfigure}
	\caption{Inter-facial shear for Rayleigh-Taylor instability as a function of curvature in various sized finite domains for (a) $A=1$ and (b) various Atwood numbers.}
	\label{fig:shearall}
\end{figure}

Solving (\ref{eq:sheardef}) to the first-order and taking the first two harmonic modes finds that the shear function is dependent on boundary size, bubble curvature and Atwood number. The explicit analytic expression is cumbersome and so not presented here, instead Figure \ref{fig:shearall} demonstrates the dependencies of the shear function. The shear function is a non-monotone function of curvature and has a maximum corresponding with the critical curvature defined previously $|\zeta_{cr}/k|= |\coth(kZ)/2|$. Truncated to the region $0\leq |\zeta_{1}| \leq |\zeta_{cr}|$ the shear function is in fact directly related and a one-to-one function of curvature. This range consists of all the possible shear function values supported by the system.\\
\\
For any fixed curvature value a decreasing domain size corresponds to an increasing shear function. RTI bubble structures evolving in a finite domain have a greater amount of shear function (and hence more distinct vortical structures) at the interface when compared to their evolution in a larger or infinite spatial domain. Irregardless of the size of the domain, instabilities growing in a highly contrasting fluid system $A\approx1$ have a larger contribution from inter-facial shear function than those with similar fluid densities $A\approx0$. Further, instabilities at all Atwood numbers converge at the same rate with increasing boundary size.\\
\\
Particular attention should be focused on the solution with maximum shear function. This solution is denoted the critical bubble with the maximum shear function value given as $\Gamma_S$ and corresponding curvature as $\zeta_S$. The critical bubble satisfies the equation
\begin{equation}
48A^4+64A^3\Gamma_S^2-24(A^2-A^4)\Gamma_S^4+(1-2A^2+A^4)\Gamma_S^8=0
\end{equation}
Solving this finds the analytic expression for the maximum shear function value and corresponding bubble front curvature respectively as 
\begin{equation}
\Gamma_S=\sqrt{\frac{2A T1 T2}{(1+A)(2T1-T2)}}(gk)^{-1/2}\quad,\quad |\zeta_S|=\frac{T1 T2 }{2(2T1-T2)}k
\end{equation} 
The dimension of the shear function, $s^{-1}$, comes from the foundational scale $(gk)^{-1/2}$. The maximum shear function value increases for contrasting fluid densities and for smaller domain sizes. The bubble curvature that maximises shear function is independent of Atwood number and is a function of the boundary size. RTI bubbles evolving in a smaller domain experience maximum inter-facial shear function for more curved bubble fronts. In a finite domain $|\zeta_S|<|\zeta_{cr}|$, whereas in the limit of an infinite domain with $kZ\rightarrow\infty$ the values are
\begin{equation}
\Gamma_S=\sqrt{\frac{2A}{1+A}}(gk)^{-1/2}\quad \text{ and } \quad|\zeta_S|=\frac{k}{2}
\end{equation} 

\subsection{Shear function as a parameter to the family of solutions}

\begin{figure}[H]
	\centering
	\includegraphics[width=\linewidth]{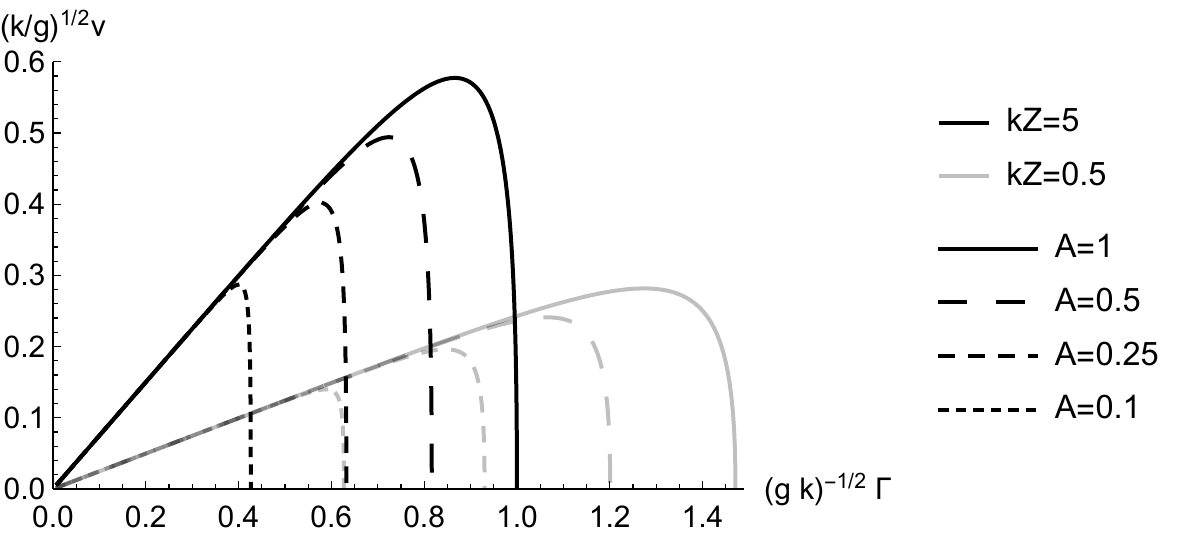}
	\caption{Velocity solutions for Rayleigh-Taylor instability as parametrised by shear in a large and small finite domain for different Atwood numbers.}
	\label{fig:vvsgamma}
\end{figure}

\begin{figure}[H]
	\centering
	\begin{subfigure}{0.49\textwidth}
		\centering
		\includegraphics[width=\linewidth]{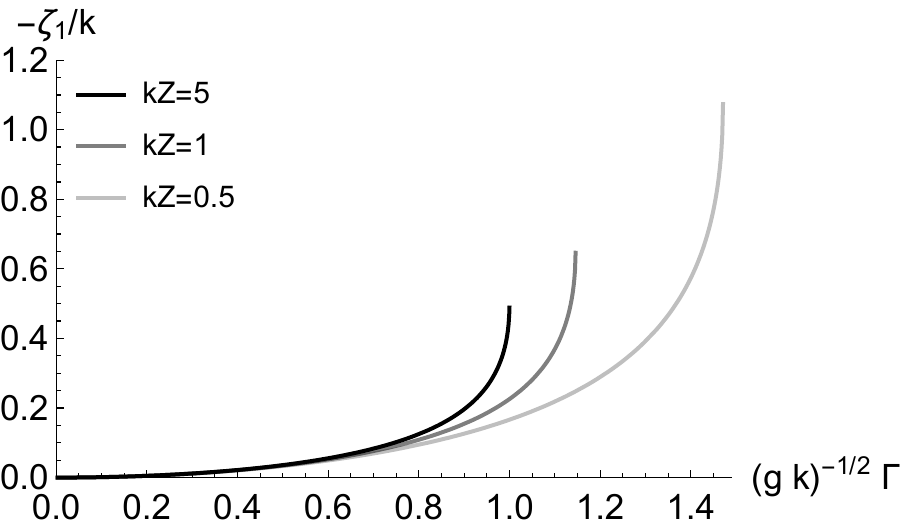}
		\caption{}
	\end{subfigure}
	\begin{subfigure}{0.49\textwidth}
		\includegraphics[width=\linewidth]{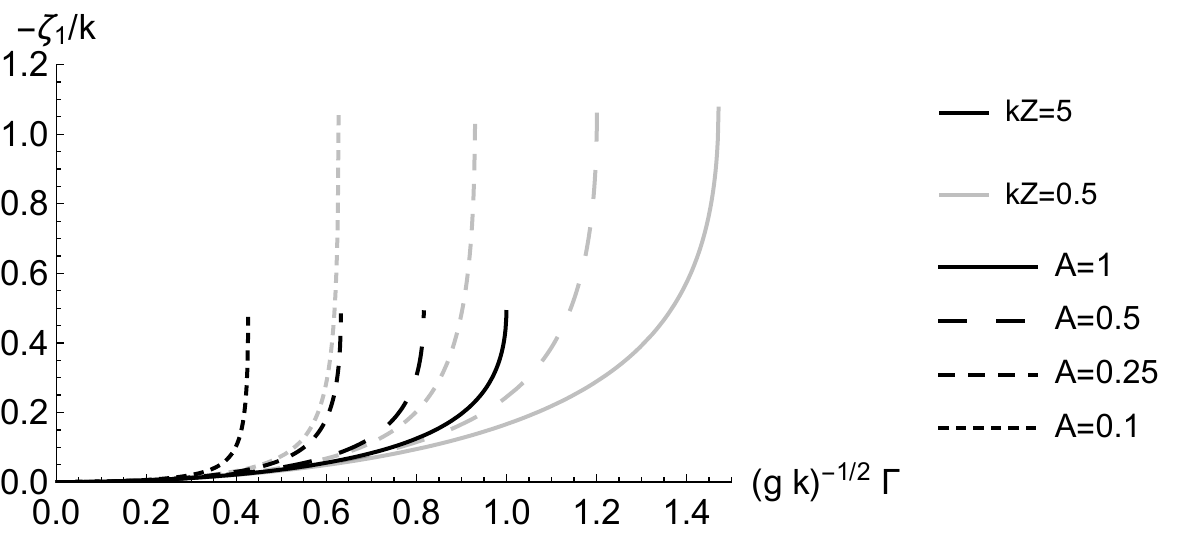}
		\caption{}
	\end{subfigure}
	\caption{Possible Rayleigh-Taylor instability bubble curvatures using inter-facial shear as the parameter to the family of solutions for (a) A=1 and (b) for fluid systems with a range of Atwood numbers.}
	\label{fig:shearvscurve}
\end{figure}

The direct link between the multiplicity of solutions and the shear function is explored by considering the shear function as an alternate parameter to the family of solutions. Figure \ref{fig:vvsgamma} illustrates the family of bubble velocity solutions in the nonlinear regime as parametrised by the shear function.\\
\\
For each Atwood number the range of velocity values is conserved whether using a curvature or a shear function parametrisation.  An interesting property of Figure \ref{fig:vvsgamma} is that the bubble velocity appears independent of Atwood number until the vicinity of maximum shear function (which is dependent on Atwood number) in which the solution falls quickly to zero. RTI evolution in a smaller finite domain is characterised by a lower maximum velocity and a greater range of possible inter-facial shear function values. \\
\\
Figure \ref{fig:shearvscurve} illustrates the family of bubble curvature solutions as parametrised by the shear function. A direct, monotone relationship of the two diagnostic parameters is observed. Using the shear function as an alternate parameter has the benefit of tying together the diagnostic parameters in the bubble front vicinity.

\section{Study of convergence}

\begin{figure}[H]
	\centering
	\begin{subfigure}{0.49\textwidth}
		\includegraphics[width=\linewidth]{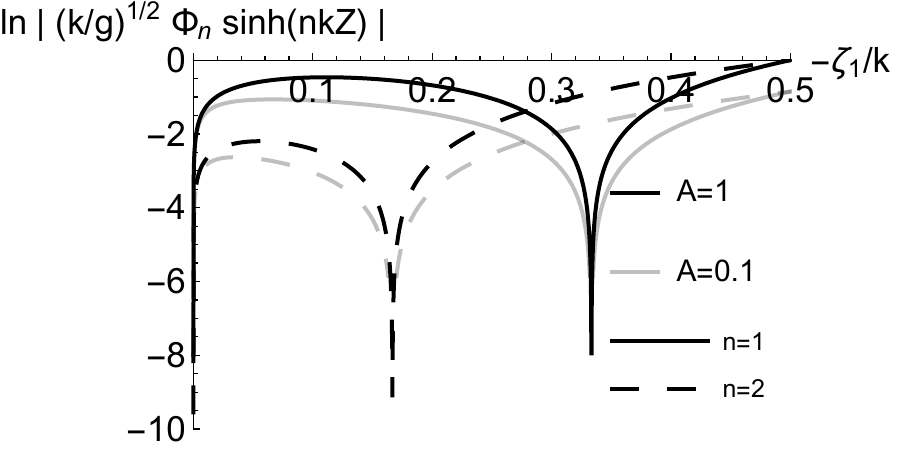}
		\caption{}
	\end{subfigure}
	\begin{subfigure}{0.49\textwidth}
		\includegraphics[width=\linewidth]{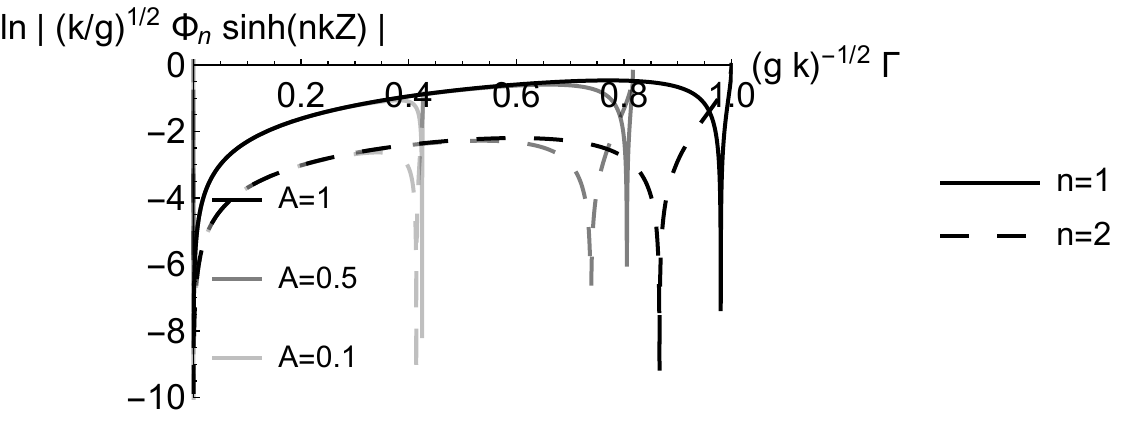}
		\caption{}
	\end{subfigure}
	\caption{Fourier amplitudes of the heavier fluid for Rayleigh-Taylor instability in a fixed finite boundary $kZ=5$ with contrasting and similar density fluids. The amplitudes are parametrised by (a) curvature and (b) shear.}
	\label{fig:convchangingAlog}
\end{figure}

The regular asymptotic solutions for the late-time evolution involve multiple harmonics and for all Atwood numbers the lowest order harmonic is dominant \cite{Inogamov}. This allows us to study the convergence properties of the solution. To prove a good convergence of the solution, it should be shown that for all harmonic modes $m$, $\Phi_{m+1} > \Phi_m$. In this analysis, in which the moments are truncated to retain two harmonic modes, the convergence of the solution is investigated by comparing the magnitude of the first, $\Phi_1$, and second, $\Phi_2$, Fourier amplitude modes. In the case that $\Phi_1>\Phi_2$ indicates that the lowest harmonic is dominant and our solution has good convergence. Both Fourier amplitudes are plotted for $0\leq|\zeta_1|\leq|\zeta_{cr}|$ and the corresponding shear function region.\\
\\  
The regions in which $\Phi_1$ represented by a continuous line lies below $\Phi_2$, the dashed line, indicates regions of poor convergence. In Figure \ref{fig:convchangingAlog} we explore the convergence properties of the heavy fluid for various Atwood numbers at a fixed boundary size $kZ=5$.\\

\section{Conclusion}

One of the fascinating things about fluid flow is the incredible richness of the mathematics and physics that appears in the problems. The challenge of describing RTI flows is no exception. We learn the importance of accurate approximations to deal with these flows. In this work, approximations such as neglecting  non-idealised fluid effects, a scale separation and approximating a potential flow enables the Navier-Stokes equations to be solved and yields a description of RTI in the vicinity of the bubble tip.\\ 
\\
Long-standing problems such as this, challenge us to find innovative techniques and approaches. The group theory approach with a strong consideration for the symmetries of the flow, is one such creative technique that helps find a description of RT evolution. The large scale bubble structure have symmetry properties defined by the $pm1$ symmetry group. \\
\\
Under the influence of a constant acceleration the initial wave perturbation curvature and growth rate increase exponentially with time. Imposing a finite domain hinders this growth by an exponent of $\tanh(kZ)$ dependent on dimensionless boundary position $kZ$.\\
\\
In the more involved nonlinear regime, solving the Navier-Stokes equations for the large-scale bubble structures with symmetry group $pm1$ finds a parameter family of solutions supported by the symmetries. The fastest solution in this family is physically significant. For RTI the fastest solution corresponds to a curved bubble. Introducing a finite domain decreases the maximum velocity and increases the possible bubble front curvatures. For completeness a description was found for all Atwood numbers, however as the scale separation breaks down for similar density fluids we should be cautious in this vicinity.\\
\\
The multiplicity of solutions in the nonlinear regime is a consequence of the governing boundary condition enabling inter-facial shear to develop. In RTI flows the shear function is a monotone increasing function of curvature. Using the shear function as an alternate parameter to the multiplicity of solutions has the benefit of tying together the observables and improving the convergence properties of the harmonic amplitudes.

\section{Further work}

The work discussed in this paper sets the groundwork for further research. Specifically, we should expand our understanding of the evolution of RTI in a finite domain by considering the three-dimensional case. A group theory approach is applicable in three dimensions and so the method of solution discussed here is appropriate with some alterations. The properties of the diagnostic parameters in the transition between three and two dimensional spatially bounded flows should also be considered. The evolution of the bubble front in a finite domain was described only to a first order approximation, $N=1$. To better the accuracy of our description of RTI evolution we should consider higher orders of approximation. \\
\\
Our suggestion that the shear function provides a better  parametrisation to the family of solutions should also be explored further. The stability of the diagnostic parameters as a function of the shear function should be found by perturbing the solutions or using bifurcation theory. It must also be checked that the dependence of the diagnostic parameters on time in the nonlinear regime remains the same in a shear function parametrisation. For example as a function of curvature, velocity is constant in time in the nonlinear regime, $v\sim O (1)$. However with a different parametrisation this should be tested accordingly. \\
\\
A closer analysis of inter-facial shear may reveal properties of the Kelvin-Helmholtz vortical structures. Specifically we may identify the influence of non-idealised properties of the fluids, such as viscous stress and surface tension, on the shear function. Numerical simulations should be run including non-idealised effects. The results provided here give a quantitative `ideal' comparison to any future numerical simulations. This may also reveal if the non-idealised effects hinder the shear function near the bubble-tip and hence explain the discrepancy of the vortical structures in the vicinity of the bubble tip.\\
\\
This paper is ended with a discussion on the possible avenues to expand this work as a motivation for future studies. This motivation stems from the need to understand and control RTI in a range of industrial applications. Although our understanding of RTI flows has greatly enhanced since Lord Rayleigh first defined the fluid problem more than $100$ years ago, there is still a need for the development of powerful theoretic approaches to facilitate a greater understanding of these fascinating flows. However, describing the influence of a finite domain and exploring the idea that using the shear function as a parameter may lead to more accurate results, brings us one step further.

\bibliographystyle{plain}
\bibliography{thesisbib} 

\clearpage

\clearpage

\clearpage

\clearpage

\clearpage

\clearpage

\clearpage

\clearpage

\clearpage

\clearpage

\clearpage

\begin{figure}[H]
	\centering
	\begin{subfigure}{0.49\textwidth}	
		\includegraphics[width=\linewidth]{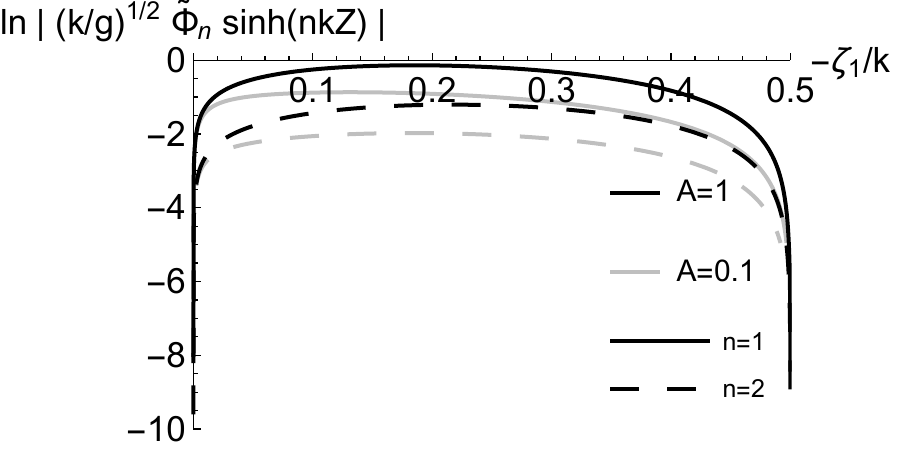}
		\caption{}
	\end{subfigure}
	\begin{subfigure}{0.49\textwidth}
		\includegraphics[width=\linewidth]{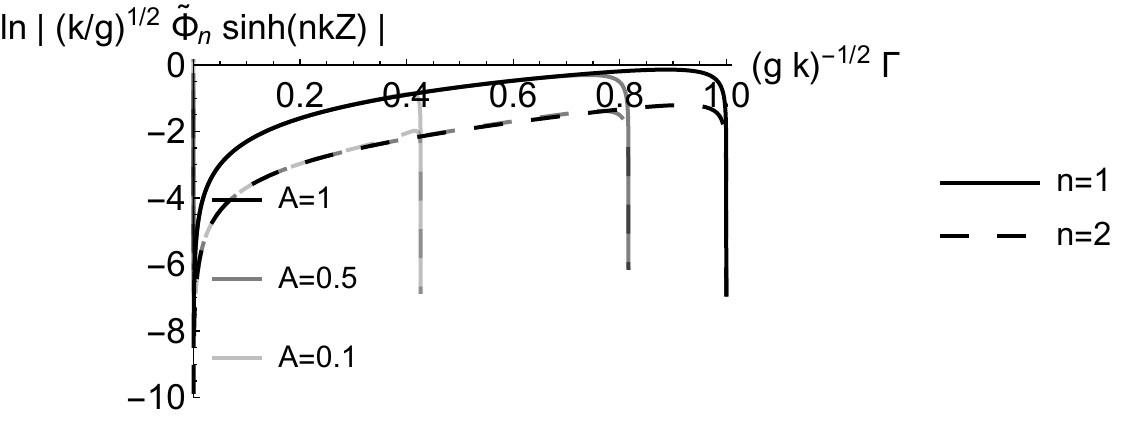}
	\end{subfigure}
	\caption{Fourier amplitudes of the light fluid of Rayleigh-Taylor instability with a fixed domain of size $kZ=5$. Similar and contrasting densities are considered. There are no issues with convergence for either parametrisation of (a) curvature or (b) shear.}
	\label{fig:convtildeAlog}
\end{figure}

\end{document}